\documentstyle[prd,aps,preprint]{revtex}
                        
\newcommand\ee{\end{equation}}
\newcommand\be{\begin{equation}}
\newcommand\eea{\end{eqnarray}}
\newcommand\bea{\begin{eqnarray}}

\newcommand\GeV{\,\mbox{GeV}}

\newcommand\mpl{M_{{\rm Pl}}}
\newcommand\Mpl{M_{\rm Pl}}

\newcommand\lsim{\mathrel{\rlap{\lower4pt\hbox{\hskip1pt$\sim$}}
    \raise1pt\hbox{$<$}}}
\newcommand\gsim{\mathrel{\rlap{\lower4pt\hbox{\hskip1pt$\sim$}}
    \raise1pt\hbox{$>$}}}

\def\dslash{\not{\hbox{\kern-2pt $\partialrtial$}}}
\def\Dslash{\not{\hbox{\kern-4pt $D$}}}
\def\Oslash{\not{\hbox{\kern-4pt $O$}}}
\def\Qslash{\not{\hbox{\kern-4pt $Q$}}}
\def\pslash{\not{\hbox{\kern-2.3pt $p$}}}
\def\kslash{\not{\hbox{\kern-2.3pt $k$}}}
\def\qslash{\not{\hbox{\kern-2.3pt $q$}}}
 \newtoks\slashfraction
 \slashfraction={.13}
 \def\slash#1{\setbox0\hbox{$ #1 $}
 \setbox0\hbox to \the\slashfraction\wd0{\hss \box0}/\box0 }
 
\def\eeq{\end{equation}}
\def\beq{\begin{equation}}

\begin{document}
\topmargin-2.5cm
%
\begin{titlepage}
\begin{flushright}
CERN-TH/99-84\\
IEM-FT-178/98\\
IFT-UAM/CSIC-99-11
\end{flushright}
\vskip 0.3in
\begin{center}
{\Large\bf 
$F$-term inflation in Superstring Theories}

\vskip .5in
{\bf J.A. Casas~$^{1,2,3,}$\footnote{\baselineskip=16pt E-mail: {\tt casas@mail.cern.ch}}},
{\bf G.B. Gelmini~$^{4,}$\footnote{\baselineskip=16pt E-mail: {\tt
gelmini@physics.ucla.edu}}} and {\bf A. Riotto~$^{3,}$\footnote{\baselineskip=16pt
 Email:
{\tt riotto@nxth04.cern.ch}}}

\vskip.35in
$^{1}$~Instituto de Estructura de la materia, CSIC, Serrano 23, 
28006 Madrid, Spain

\vskip 0.3cm

$^{2}$~Instituto de F\'{\i}sica Te\'orica, Univ. Aut\'onoma de Madrid,
28049 Madrid, Spain

\vskip 0.3cm

$^{3}$~Theory Division, CERN, CH-1211 Geneva 23, Switzerland

\vskip 0.3cm

$^{4}$~ Department of Physics and Astronomy, UCLA, Los Angeles, CA 90095-1547

\end{center}

\vskip1.3cm
\begin{center}
{\bf Abstract}
\end{center}
\begin{quote}

A supersymmetric inflationary stage dominated by an $F$-term  has the
problem  that the flatness of the potential is spoiled by supergravity
corrections, that is the slow-roll parameter $\eta$ gets
contributions of order unity.  We show that in $F$-term inflationary
models based on strings there is  natural way of obtaining small
values of $\eta$. This happens in models of hybrid inflation based on
orbifold constructions, in which a modulus $T$ field is responsible
for the large value of the potential  during inflation, and a second
field $\phi$ with appropriate modular  weight is responsible for the
roll-over. We illustrate the mechanism with a model in which the
inflaton potential is provided by gaugino condensation, leading to
succesful inflation.

\end{quote}

\vspace{2cm}

{\large March, 1999}

\vskip1.cm
\end{titlepage}

\setcounter{footnote}{1}
\setcounter{page}{1}
\newpage
%
\baselineskip=20pt

\noindent


The flatness and the horizon problems of the standard big bang
cosmology are elegantly solved if during the evolution of the early
Universe the energy density happens to be dominated by some form of
vacuum energy and comoving scales grow quasi-exponentially
\cite{guth}.  An inflationary stage is also required to dilute any
undesirable topological defects left as remnants after some phase
transition taking place at early epochs. The vacuum energy driving
inflation is generally assumed to be associated to some scalar field
$\phi$, the inflaton, which is initially displaced from the minimum of
its potential. As a by-product, quantum fluctuations of the inflaton
field may be the seeds for the generation of structure. The levels of
density and temperature fluctuations observed in the present Universe,
$\delta\rho/\rho\sim 10^{-5}$, require the inflaton potential to be
extremely flat. Actually, for inflation to be viable, there are two
parameters which must be small with respect to unity to insure
sufficient  slow roll in the inflationary potential $V$. These are the
so called  $\epsilon=\frac{1}{2}\Mpl^2(V'/V)^2$ and  $\eta=\Mpl^2
V''/V$ parameters.  Here $\Mpl=2.4\times 10^{18}$ GeV is the reduced
Planck mass. The value of $\eta$ in particular is constrained by the
observational bound in the spectral index $|n-1|\lsim 0.2$. Since in
an inflationary scenario $n=1-6\epsilon+2\eta$, this requires
$|\eta|\lsim 0.1$ (the contribution proportional to $\epsilon$ is
usually negligible ). Note that $\eta$ is essentially given
by the effective inflaton mass during inflation.

The required extreme flatness of the inflaton potential is the main
reason why inflation is more natural in the context of supersymmetric
(SUSY)  theories. Flat directions are frequent in supersymmetric
theories, and they are stable under radiative corrections, as long as
SUSY is not broken.  However, the previous argument does not hold if
SUSY is broken, which necessarily occurs during inflation, when there
is a vacuum energy density $\langle V \rangle_{{\rm in}} \sim H^2
\Mpl^2 >0$. Thus, one generically expects effective soft terms, in
particular soft scalar masses, of order  $H$, spoiling the desired
flatness of the potential. 

  The supergravity potential $V$ consists of two pieces --- the
so-called $D$-term and $F$-term. The latter is given by 
\be
V_F= e^{G}\left( G_{\bar j} K^{\bar j l}G_{l}-3 \Mpl^4\right)
= F^{\bar l} K_{j \bar l}F^j-3e^{G} \Mpl^4~.
\label{VFD}
\ee
Here $G=K \Mpl^{-2}+\ln(|W|^2 \Mpl^{-6})$, where $W$ is the
superpotential, $K^{\bar l j}$ is the inverse of the K{\"a}hler metric
$K_{j \bar l}\equiv \partial K /\partial \phi_j \partial \bar \phi_l$,
$\phi_j$  are the (scalar components) of the chiral superfields and
$F^j=e^{G/2 }K^{\bar k j}G_{\bar k}\Mpl^3$ are the corresponding
auxiliary fields.  In the following, we will assume that during
inflation the potential is dominated by the $F$-term, i.e. $V\simeq
V_F$, which is the so-called $F$--inflation.  Notice that the $F$-term
depends only upon two objects --- the superpotential   $W$, an
holomorphic function of the scalar fields,  and the  K\"{a}hler
potential  $K$, a real function.  During inflation
\be
\langle V \rangle_{{\rm in}} \equiv V_0 = 3H^2 \Mpl^2,
\label{Vi}
\ee
which implies that some $F$ fields are different form zero, thus
breaking supersymmetry. 

It is easy to check that this breaking of SUSY generically gives a
sizeable soft mass-squared to the inflaton. 
Using a basis where the fields have
canonical normalization at the origin ($K_{i \bar j}=\delta_{ij}$ at
$\phi_i=0$), a generic scalar mass 
$m_i^2 =\partial^2V/\partial \phi_i\partial \phi_i^*$ is given by
\begin{eqnarray}
 m_i^2 = \mpl^{-2}V_0 + e^K \sum_{nm}K^{n ^*m}_{i i^*} W_n W_m^* 
-\mpl^{-2} e^K |W_i|^2 + e^K\sum_n|W_{ni}|^2 +\cdots. 
\label{mphi98}
\end{eqnarray}
%
The contribution of the first term to  $\eta\equiv
\mpl^{-2}m_\phi^2V^{-1}$ is precisely 1. For generic models, the
contribution of the second term to $\eta$ is also ${\cal O}(1)$. So, even if
in the global SUSY limit the inflaton is not responsible for the SUSY
breaking ($W_\phi=0$), remaining as a flat direction during inflation
($W_{n\phi}=0$), we see from eq.(\ref{mphi98}) that the mass-squared
of the inflaton is expected to be ${\cal O}(\mpl^{-2}V)$, and thus
$\eta={\cal O}(1)$.  Therefore, it is generally   difficult to naturally
implement a slow-roll inflation in the context of supergravity (recall
we need $|\eta|\lsim 0.1$).

 Solutions to the $\eta$-problem already exist in the literature
\cite{lr}. Among them, $D$-term inflation seems to be a promising one
\cite{dterm}. In this paper, however, we show that in $F$-term
inflationary models based on (weakly coupled) strings there is a
natural way of obtaining small values of $\eta$. This happens in
models of hybrid inflation\cite{Linde} based on orbifold
constructions, in which a modulus $T$ field is responsible for the
large value of the potential $V$ during inflation, and a second field
$\phi$ with modular weight $n_\phi=-3$ is responsible for the rollover
\cite{cg}.

String models of  $F$-term inflation scenarios  present special
characteristics because the K{\"a}hler potential, $K$, which plays
a crucial role, is greatly constrained. 
In orbifold constructions,
the tree--level  K{\"a}hler potential is given by \cite{DKL}
\be
K =  -\log(S+\bar S)-3\log(T+\bar T)+\sum_j(T+\bar T)^{n_j}|\phi_j|^2 +
{\cal O}\left(\frac{|\phi_j|^4}{\Mpl^{-2}}\right))~.
\label{Kor}
\ee
Here $S$ is the dilaton and $T$ denotes generically the moduli 
fields, all in units of $\Mpl$, $\phi_j$ are the chiral
fields and $n_j$ the corresponding modular weights. The latter depend
on the type of orbifold considered and the twisted sector to which
the field belongs. The possible values of $n_j$ for $Z_N$ orbifolds are 
$n_j= -1,-2,-3,-4,-5$. The discrete character of $n_j$ will play
a relevant role later on.
  Eq.(\ref{Kor}) is written with the usual simplification
of considering a single ``overall modulus'' $T$. 
$\langle S\rangle$ and $\langle T\rangle$ are expected
to have determined values of  $ {\cal O}(1)$ in Planck units
at low energy.  This comes from the fact that 
$\langle S\rangle$ and $\langle T\rangle$ have precise physical
meanings. Namely, $\langle S\rangle$ is the value of the unified
gauge coupling constant and $\langle T\rangle$ is the squared
radius of the compactified space, both in Planck units.

{}From eq.(\ref{Kor}) it is easy to understand that $T$ is a good
candidate for providing the vacuum energy during inflation (i.e.
$F^T\neq 0$). The reason is the coupling between $\phi$ and $T$
in the quadratic term ($\propto |\phi|^2$) of $K$. Notice that, without
this coupling, if $\phi$ corresponds to a flat 
direction in global SUSY breaking ($W_\phi=W_{n\phi}=0$), then $m_\phi^2=V_0\Mpl^{-2}$
(see eq.(\ref{mphi98})) and thus $\eta=1$. 

If SUSY is broken by the $T$ (and/or $S$) fields, the corresponding
soft terms, in particular mass terms, for $\phi$ are straightforwardly
extracted from Eqs.(\ref{VFD}) and (\ref{Kor}). This was precisely the
sort of scenario considered in Refs. \cite{KL2,BIM}.  Although the
motivation of these works was different, namely to study the form of
the soft breaking terms at low--energy  with $m_{3/2}={\cal O}(1$
TeV),  their results are applicable here. The only difference here is
that the scale of the breaking is much higher and the non-vanishing
cosmological constant, $V_0>0$, plays a major role. In particular, the
effective $\phi$  mass squared, $m_\phi^2$, of a chiral field $\phi$
with modular weight  $n_\phi$, for the canonically normalized field
$(K_{\phi\bar\phi})^{1/2}\phi$, is given by \cite{BIM}
\be
m_\phi^2 = m_{3/2}^2\left\{(3+n_\phi\cos^2\theta)C^2-2\right\}~,
\label{mphi4}
\ee
where the effective gravitino mass during inflation, $m_{3/2}$, 
is defined by
\be
m_{3/2}^2=e^{G} \Mpl^2= e^{K/\Mpl^2}|W|^2 \Mpl^{-4}~. 
\label{mgravitino}
\ee
(since $V_0=F^{\bar l} K_{j \bar l}F^j-3m_{3/2}^2 \Mpl^2$,
unless there is some fine--tuning, $m_{3/2}^2$ is at most  
${\cal O}(V_0) \Mpl^{-2}$); 
$\tan^2\theta= (K_{S \bar S}/
K_{T \bar T})\left|F^S / F^T\right|^2$ and 
$C^2= 1+ [V_0 /(3\Mpl^2m_{3/2}^2)]$. Notice that $C^2>1$.

{}From (\ref{mphi4}) we conclude the following.  If $\cos^2\theta=0$,
the case of $S$--driven inflation, then $m_\phi^2 =m_{3/2}^2+V_0
\Mpl^{-2}$. Hence, $\eta={\cal O}(1)$
as expected from previous arguments. Let us examine
now the more promising case of $T$-driven inflation, where $\cos^2\theta=1$. 
Note that in this case $V_0= K_{T \bar T}|F^T|^2-3m_{3/2}^2 \Mpl^2=3H^2
\Mpl^2$.
If it happens that $K_{T \bar T}|F^T|^2 = {\cal O}(m_{3/2}^2) \Mpl^2$, what
means $C^2={\cal O}(1)$, then still one obtains the usual result of
$m_\phi^2 = {\cal O}(m_{3/2}^2)= {\cal O}(H^2)$, and thus $\eta={\cal O}(1)$.
However, it may
perfectly happen that 
\be
K_{T \bar T}|F^T|^2\gg
m_{3/2}^2 \Mpl^2. 
\label{ineq1}
\ee
Then $V_0 \gg {\cal O}(m_{3/2}^2 \Mpl^2)$ and
$C^2\gg 1$. Then Eq. (\ref{mphi4}) simplifies to
\be
m_\phi^2\simeq \frac{1}{3}V_0\Mpl^{-2}(3+n_\phi)=H^2(3+n_\phi). 
\label{msmall}
\ee
Hence, for $n_\phi=-3$ we get $m_\phi^2\ll {\cal O}(H^2)$ and thus 
$\eta\ll 1$, as desired\footnote{To get this result we have assumed
eq.(\ref{ineq1}), which is equivalent to 
$|W_T|\gg\frac{9}{(T +\bar T)^2}|W|^2$. This is easily
fulfilled for slightly large values of $T$. E.g. if $W\propto
[\eta(T)]^{-6}$, as dictated by target-space modular invariance, 
then $|W_T|\gg\frac{5}{2}|W|^2$ and $|\eta|<0.1$ for Re$T>2.5$.}.

Therefore, we have shown that the possibility of a very small mass
$m_\phi^2$  can occur in $T$ driven $F$-term inflation, if $V_0 \gg
{\cal O}(m_{3/2}^2 \Mpl^2)$ (namely $C^2\gg 1$) and the modular weight
$\phi$ is  $n_\phi= -3$. Notice that this possibility appears with no
fine tunning at all, thanks to the discrete character of $n_\phi$,
thus avoiding the slow-roll $\eta$-problem. In this scenario the
energy density of the vacuum is provided by the modulus $T$ and the
slow-roll field is the $\phi$ field. We are envisaging, therefore, an
hybrid inflationary scenario where the dynamics of the system will
make the energy density of the vacuum disappear for some critical
value of the field $\phi$. 

To get these nice results we have used the form of the K{\"a}hler
potential of eq.(\ref{Kor}), which contains two main approximations:
{\em (i)} it is a tree-level expression, and {\em (ii)} it is leading
order in the expansion in powers of $|\phi|^2$.

Concerning approximation {\em (i)}, perturbative corrections to
eq.(\ref{Kor}) are known at one-loop level \cite{DKL} and are small,
so they do not affect any of the results presented here.  More
precisely, at one loop the correction consists in  replacing in
Eq. (\ref{Kor}) 
\be 
\label{oneloop}
S+{\bar S} \to Y= S +{\bar S} - { \delta_{GS}
\over 8 \pi^2} {\rm ln}(T+{\bar T}), 
\ee 
where $\delta_{GS}$ is  the
Green-Schwarz parameter, which is negative and $|\delta_{GS}|  \le
{\cal O}(10)$. With this replacement, Eq. (\ref{mphi4}) becomes
\be
m_\phi^2 = m_{3/2}^2\left\{\left[3+n_\phi
\left(1-{ \delta_{GS} \over 24 \pi^2 Y}\right)^{-1}   
\cos^2\theta\right]C^2-2\right\}~.
\label{mphiGS}
\ee
In the case we identified above, in which $\cos^2 \theta =1$, $C^2 \gg 1$
and $n_\phi = -3$, from (\ref{mphiGS}) we obtain
\be
m_\phi^2\simeq - H^2 {\delta_{GS} \over 8 \pi^2 Y} 
\simeq - H^2 {\delta_{GS} \over 32 \pi^2 }, 
\label{msmallGS}
\ee
where we have used  that $Y$ is the inverse of the unified coupling
constant $g$, namely $Y\simeq 2/g^2$, so we expect $\langle Y\rangle
\simeq 4$.  Eq.(\ref{msmallGS}) shows that $m_\phi^2 < 0.1~ H^2$,
which leads to  values of $\eta$ small enough to allow for the
implementation of inflation. On the other hand, non-perturbative corrections 
are very poorly known (see e.g. Ref. \cite{Ca} for an analysis of their 
possible phenomenological significance).

Concerning approximation {\em (ii)}, the K{\"a}hler potential will
generically contain terms of the form
\be
K= K_0 + K_{\phi\bar\phi}|\phi|^2 + \lambda_4|\phi|^4+ 
\lambda_6 |\phi|^6+ ... ,
\label{Kexpand}
\ee
where the first two terms are the ones explicited in eq.(\ref{Kor}),
and one expects $\lambda_n={\cal O}(\Mpl^{2-n})$. These extra terms modify
the potential to be
\be
V= V_0 \left(1 + \kappa_1^4 |\phi|^4+ \kappa_2^6 |\phi|^6+ ...\right) ~,
\label{Vexpand}
\ee
where the coefficients $\kappa_i$ are all of the order of $\Mpl^{-1}$.
For $\phi \lsim 0.1 \Mpl$ the $|\phi|^4, |\phi|^6, ...$ terms become
negligible (this is a common situation and, as we will see, it is the 
case of the particular model we present below). For example, keeping only
the $|\phi|^4$ term above, and  ignoring the phase of $\phi$,  
the $\epsilon$ and $\eta$ parameters are given by 
\be
\epsilon= { 8 \Mpl^2 \kappa_1^8 \phi^6 \over (1+ \kappa^4 \phi^4)^2}~,
\;\;\;\;
\eta= {12 \Mpl^2 \kappa_1^2 \phi^2 \over 1 + \kappa_1^2 \phi^4}.
\label{eps}
\ee
Then, if $\phi \lsim 0.1 \Mpl$ we get 
$\epsilon \simeq 8 \kappa_1^6 \phi^6 \lsim
10^{-5}$, $\eta \simeq 12 (\phi/\Mpl)^2\lesssim 0.1$.

It is amusing to note that the exact (tree-level) form of $K(\phi, \bar\phi)$
is not known, but has been conjectured \cite{ferrara}. For our case of 
interest,
$n_\phi=-3$, it simply becomes
\be
K =  -\log(S+\bar S)-\log\left[(T+\bar T)^3 - |\phi|^2\right].
\label{Kor2}
\ee
Then, assuming $V_0\gg\Mpl^2 m_{3/2}^2$ (i.e. $C^2\gg 1$), it turns
out that all the $|\phi|^4, |\phi|^6, ...$ contributions in 
(\ref{Vexpand}) are exactly vanishing!

We have discussed a natural way to solve the slow-roll problem (i.e. to get
$|\eta|\ll 1$ during inflation). However, 
without knowing the $T$-dependent potential we cannot know the value
of the field $\phi$ at the end of inflation $\phi_e$. 
Neither we can calculate  
its value 60 inflation-folds before the end, $\phi_{60}$. This is important
to get more definite results. Next we show a model based on gaugino 
condensation in which the previous mechanism to get $|\eta|\ll 1$ is 
naturally implemented, allowing for further predictions.

Suppose the hidden sector of the theory has an $SU(N_c)$ gauge group with
$N_f$ flavours. Then the gaugino condensation effects generate a 
non-perturbative superpotential $W_{np}$ of the form \cite{gaugino}
\be
W_{np}= \det {\cal M}^{1/N_c}\ f(T).
\label{Wnp}
\ee
Here ${\cal M}$ is the effective ``quark'' mass matrix and 
$f(T)\sim e^{-8\pi^2/N_cg^2}[\eta(T)]^\alpha$, where 
$\alpha={\cal O}(1)$.
Suppose that ${\cal M}$ has an invariant contribution and
a field-dependent contribution, i.e.
\be
{\cal M}= M - \lambda A^q\ ,
\label{Mcal}
\ee
where $M$ is a constant mass term (typically $M={\cal O}(\mpl)$), $A$ is 
a scalar field, $\lambda$ is a coupling constant and $q$ is a numerical 
exponent ($q=1$ or $q=2$ is okay for our purposes). 
The $A$ field can also have perturbative interactions with the 
inflaton $\phi$ (i.e. a
field with $n_\phi=-3$), in particular $\sim A^2 \phi^r$, 
where $r$ is an undetermined exponent.
Then the relevant superpotential gets the form
\be
W_{np}= \left[M - \lambda A^q\right]^{N_f/N_c}\ f(T)\ +\ \lambda' A^2 \phi^r\ , 
\label{Wtot}
\ee
where $\lambda'$ is a coupling constant. As we will see, we need
$r\geq 2$ for our model to work.

It is easy to check that the supergravity potential, $V_{SUGRA}$, 
derived from eq.(\ref{Wtot}) has a {\em global minimum} at
\be
\lambda A^q = M\;\;\;\;\;\;\;\phi=0\ .
\label{globalmin}
\ee
Notice that at this point $\det {\cal M}=0$ and $V=0$, and supersymmetry
is unbroken since
\be
W=W_T=W_A=W_\phi=0\ .
\label{SUSYunb}
\ee
On the other hand, for large $\langle\phi\rangle$ the $A$ field gets an
effective large mass and, hence, the {\em effective minimum} 
of $V_{SUGRA}$  is at
\be
A \simeq 0\ .
\label{effmin}
\ee
So, the $\phi$ potential in this region is simply given by
\be
V(\phi)=K_{T\bar T}|F^T|^2\equiv V_0\ ,
\label{flat}
\ee
which is flat, leading to an inflationary process. Actually, the
potential is not exactly flat. As explained above, perturbative 
contributions to the K{\"a}hler potential give to $\phi$ a positive 
tiny mass $m_\phi^2\simeq  H^2 |{\delta_{GS}| \over 32 \pi^2 }$ 
(see eqs.(\ref{oneloop}--\ref{msmallGS})). So, the inflaton
$\phi$ rolls slowly towards the origin.
Inflation
ends when the effective $A$ mass becomes unimportant. This occurs for
a value of $\phi=\phi_e$ given by
\be
\phi_e\simeq 
V_0^{1/2r}{\mpl}^{1-2/r}\ .
\label{phie2}
\ee
For $\phi<\phi_e$, $\phi$ and $A$ move quickly towards the true (global) minimum
and inflation ends. The value of $\phi$ 60 inflation-folds before the end, 
$\phi_{60}$, is given by
\be
\phi_{60}\simeq e^{|\delta_{GS}|/16}\phi_e\simeq 2\phi_e\ .
\label{phi60}
\ee
This is crucial to evaluate the primordial spectrum of scalar perturbations
\be
{\cal P}_\xi=\frac{\kappa^6}{3(2\pi)^2}\frac{V_{60}^3}{(V'_{60})^2}\ .
\label{pert}
\ee
According to observations ${\cal P}_\xi\simeq (5\times 10^{-5})^2$,
requiring
\be
V_0^{1/4}=\left[2\times 10^{-5}\right]^{\frac{r}{2(r-1)}}\mpl\ .
\label{V0}
\ee
More precisely, for $r\geq 2$ this translates into
\be
5\times 10^{13}{\rm GeV}\leq V_0^{1/4}\leq
10^{16}{\rm GeV}\ ,
\label{V02}
\ee
where the lower bound corresponds to the simplest case $r=2$.  We
finally note that these numbers for $V_0$ are perfectly natural in our
model. More precisely, $V_0$ is given by
eqs.(\ref{flat},\ref{Wtot},\ref{effmin}), so typically is
$V_0^{1/4}\sim e^{-4\pi^2/N_cg^2}{\cal O}(\mpl)$. Thus, the 
suppression needed to bring $V_0^{1/4}$ into the range of eq.(\ref{V02})
is provided by $N_c\sim 8$, a perfectly reasonable number.
It is also important to keep in mind that the gaugino
condensate we have used for inflation does not play any role for
low-energy SUSY breaking, as it becomes vanishing at the final
(global) minimum. Hence, the magnitud of $W$ (and thus of $N_c$)
during inflation is not constrained by any low-energy SUSY breaking
requirement, which would demand $W={\cal O}(1\ TeV)\mpl^2$.

In conclusion, we have shown how the slow-roll problem, which
generically affects all the inflationary models based on supergravity
when the vacuum is dominated by an $F$-term, is naturally solved in
the context of SUGRA theories coming form superstrings. In particular,
this occurs if the vacuum energy during inflation is dominated by a
modulus ($T$) $F$-term and the role of the inflaton is played by any
field with the appropriate modular weight, namely $n=-3$. In this
case, the inflaton mass is exactly vanishing at tree-level, and gets
non-dangerous contributions at higher radiative levels. Inflation
takes place in a hibrid-like manner. We have explicitely shown that it
is easy to construct models implementing this idea successfully. 
In particular we
have presented a model based on the vacuum energy created by a gaugino
condensate in the hidden sector, coupled to the inflaton field in a
suitable way.

\vspace{0.3cm}


\vskip 0.2cm {\bf Acknowledgements}

This research was supported in part by the CICYT
(contract AEN95-0195) and the European Union
(contract CHRX-CT92-0004) (JAC), and 
by the US Department of Energy
under grant DE-FG03-91ER40662 Task C (GG).

\vskip 0.2cm
\def\MPL #1 #2 #3 {{\em Mod.~Phys.~Lett.}~{\bf#1}\ (#2) #3 }
\def\NPB #1 #2 #3 {{\em Nucl.~Phys.}~{\bf B#1}\ (#2) #3 }
\def\PLB #1 #2 #3 {{\em Phys.~Lett.}~{\bf B#1}\ (#2) #3 }
\def\PR  #1 #2 #3 {{\em Phys.~Rep.}~{\bf#1}\ (#2) #3 }
\def\PRD #1 #2 #3 {{\em Phys.~Rev.}~{\bf D#1}\ (#2) #3 }
\def\PRL #1 #2 #3 {{\em Phys.~Rev.~Lett.}~{\bf#1}\ (#2) #3 }
\def\PTP #1 #2 #3 {{\em Prog.~Theor.~Phys.}~{\bf#1}\ (#2) #3 }
\def\RMP #1 #2 #3 {{\em Rev.~Mod.~Phys.}~{\bf#1}\ (#2) #3 }
\def\ZPC #1 #2 #3 {{\em Z.~Phys.}~{\bf C#1}\ (#2) #3 }


\begin{thebibliography}{99}

\bibitem{guth} A.H. Guth, Phys. Rev. {\bf D23}, 347 (1981).  See also
A.D. Linde, {\it Particle Physics and Inflationary Cosmology}, (Harwood
Academic, New York, 1990).
%
\bibitem{grisaru79} M. Grisaru, W. Siegel nd M. Rocek, Nucl. Phys. {\bf
B159}, 429 (1979). 
%
\bibitem{f} M. Dine, W. Fischler and D. Nemeschansky,
        Phys. Lett. {\bf 136B}, 169 (1984);  G. D. Coughlan, R. Holman, P. Ramond and G. G. Ross,  Phys. Lett. {\bf 140B}, 44 (1984); E. J. Copeland, A. R. Liddle, D. H. Lyth, E. D. Stewart
        and D. Wands, Phys. Rev. D {\bf 49}, 6410  (1994).
%
\bibitem{lr} D. H. Lyth and A. Riotto, {\it Particle Physics Models of inflation and the cosmological  density perturbation}, hep-ph/9807278,
to appear in  Phys. Rept. 
%
\bibitem{dterm}   
J.A. Casas and C. Mu{\~n}oz, \PLB 216 1989 37;
J.A. Casas, J.M. Moreno, C. Mu{\~n}oz and M. Quir\'os, 
\NPB 328 1989 272;
P. Binetruy and G. Dvali, Phys. Lett. {\bf B388}, 241
(1996); E. Halyo, Phys. Lett. {\bf B387}, 43 (1996); D.H. Lyth and A. Riotto, hep-ph/9707273, Phys. Lett. {\bf
412}, 28 (1997); G. Dvali and A. Riotto, hep-ph/9706408,  Phys. Lett. {\bf B417}, 20 (1998); C. Kolda and J. March-Russell, hep-ph/9802358;
J.R. Espinosa, A. Riotto and G.G. Ross, CERN-TH-97-7, hep-ph/9804214, Nucl. Phys. {\bf B531}, 461 (1998). 
%
\bibitem{Linde} A. Linde, \PLB 259 1991 38; \PRD 49 1994 748.
%
\bibitem{cg}
J. A. Casas and G. B. Gelmini,  Phys. Lett. {\bf B410} (1997) 36.
%
\bibitem{DKL} L. Dixon, V. Kaplunovsky and J. Louis, 
\NPB 329 1990 27; J.--P. Derendinger, S. Ferrara, C. Kounnas
and F. Zwirner, \NPB 372 1992 145.
%
%
%
%
\bibitem{KL2} V. Kaplunovsky and J. Louis,  \PLB 306 1993 269.
%
\bibitem{BIM} A. Brignole, L.E. Ib{\'a}{\~n}ez and C. Mu{\~n}oz, 
\NPB 422 1994 125.
%
\bibitem{Ca} J.A. Casas, \PLB 384 1996 103;
P. Binetruy, M.K. Gaillard and Y. Wu, \NPB 481 1996 109.
%
\bibitem{ferrara} S. Ferrara, D. L\"ust and S. Theisen, \PLB 233 1989 147.

\bibitem{gaugino} D. L\"ust and T.R. Taylor, \PLB 253 1991 335;
B. de Carlos, J.A. Casas and C. Mu\~noz, \PLB 263 1991 248;
D. L\"ust and C. Mu\~noz, \PLB 279 1992 272.

\end{thebibliography}
\end{document}